\pdfoutput=1
\documentclass[aps,10pt,twocolumn,floats,floatfix,showkeys,showpacs,superscriptaddress]{revtex4}
\usepackage{graphicx,amssymb,amsmath}
\usepackage[latin1]{inputenc}
\usepackage{color}
\begin{document}

\title{Understanding Information Transmission in Complex Networks}
\author{Nicol{\'a}s Rubido}
\affiliation{Universidad de la Rep\'{u}blica, Instituto de F\'{i}sica de la Facultad de Ciencias, Igu\'{a} 4225, Montevideo, 11400, Uruguay}
\author{Celso Grebogi}
\affiliation{University of Aberdeen, King's College, Institute for Complex Systems and Mathematical Biology, AB24 3UE Aberdeen, UK}
\author{Murilo S. Baptista}
\affiliation{University of Aberdeen, King's College, Institute for Complex Systems and Mathematical Biology, AB24 3UE Aberdeen, UK}
\date{\today}
\begin{abstract}
Information Theory concepts and methodologies conform the background of how communication systems are studied and understood. They are mainly focused on the source-channel-receiver problem and on the asymptotic limits of accuracy and communication rates, which are the classical problems studied by Shannon. However, the impact of Information Theory on networks (acting as the channel) is just starting. Here, we present an approach to understand how information flows in any connected complex network. Our approach is based on defining linear conservative flows that travel through the network from source to receiver. This framework allows us to have an analytical description of the problem and also linking the topological invariants of the network, such as the node degree, with the information flow. In particular, our approach is able to deal with information transmission in modular networks (networks containing community structures) or multiplex networks (networks with multiple layers), which are nowadays of paramount importance.
\end{abstract}
\keywords{Complex Networks, Flow Networks, Information Measures.}
\pacs{89.75.Hc,45.30.+s,02.50.-r,41.20-q}

\maketitle
\section{Introduction}
The Physical Universe is ruled by Physical Laws. These laws are constructed under the principle that any two bodies interact among themselves by the force between them. The interaction level is measured by the force magnitude. In the field of complex systems, an extension of this concept, and widely employed quantity to measure the interaction level between two or more systems, is their synchronisation degree \cite{Pikovsky_2003,Strogatz_2003}. Namely, a measure that quantifies the similarity between their behaviour. From the information theory perspective, similar behaviours correspond to significant information sharing, since knowing what one system is doing allows one to predict with high-accuracy what the other system is doing \cite{Shannon}. For example, the Mutual Information Rate was recently proposed as a way to measure how two dynamical systems (or group of systems) coupled in a network are related to each other \cite{Baptista1,Baptista2}, which has also proven useful to infer their structural connection \cite{Ezequiel_2016}, contrary to the use of synchronisation degree. Consequently, the main question that we raise is to understand how strong will two systems or networks interact by analysing them as a communication system.

On the one hand, using dynamical networks to generate encodings, such that the information transmission is enabled, has proven to be an important task in recent years. In this sense, an outstanding achievement is the Computing Reservoir technique \cite{Jaeger_2004,Hermans_2010,Paquot_2012}, which is somewhat an extension of the well-known neural networks. Nevertheless, the information generation/transmission underlying mechanisms and practical capabilities of such techniques are not fully understood yet. On the other hand, networks topological structures have recently been used to asses their vulnerability to cascade failures \cite{Simonsen_2008,Lai_2009,Havlin_2010,Scala_2014} and information capabilities \cite{Sole_2004}. These works discard physical quantities that might be transported by the network and focus solely in the structure, hence, they are unsuitable to be analysed as a communication system.

\begin{figure}[htbp]
 \begin{center}
  \includegraphics[width=0.8\columnwidth]{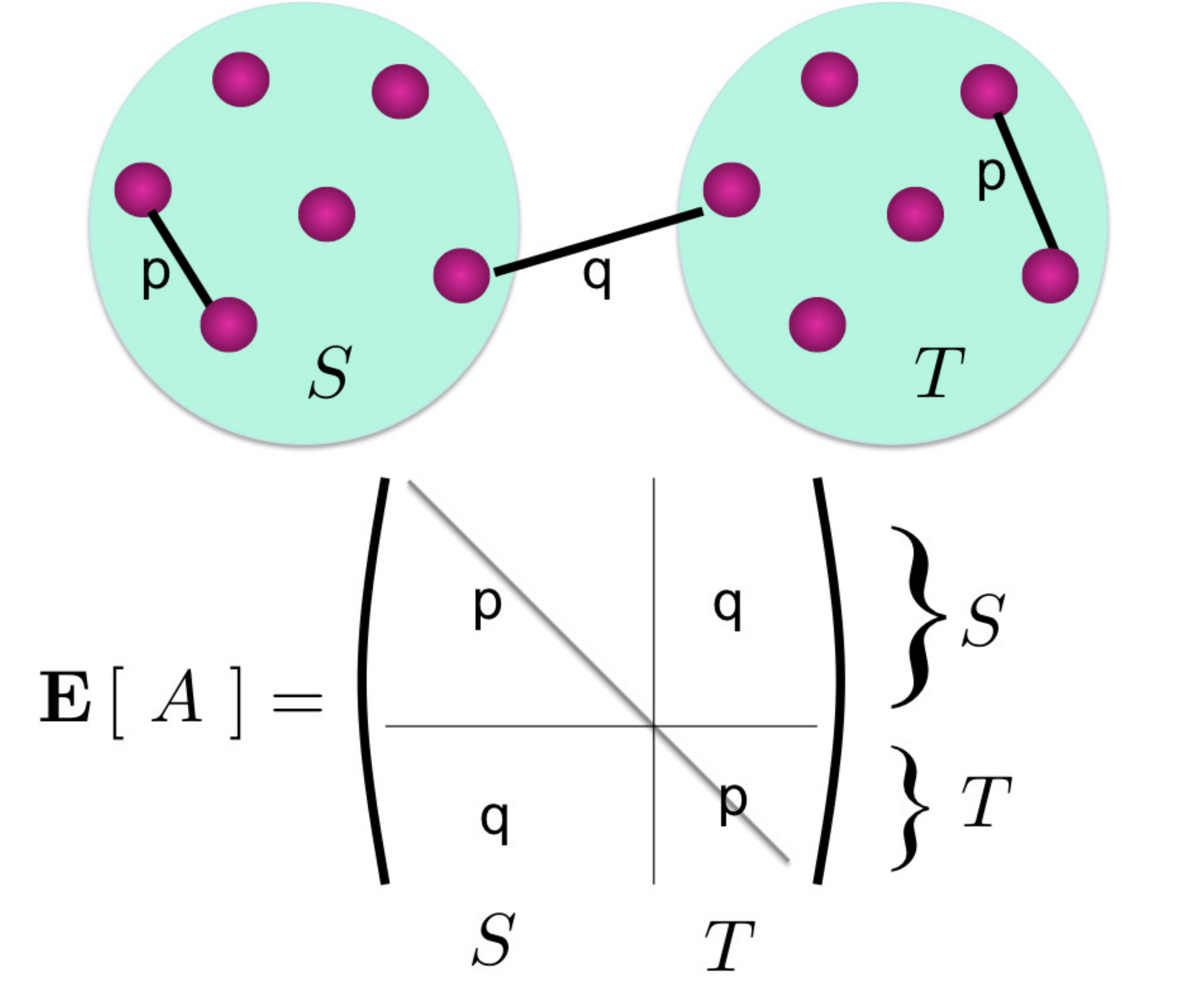}
 \end{center} \vspace{-1pc}
  \caption{The top panel shows schematically our communication network, where a division between transmitter nodes, $S$, receiver nodes $T$, and connecting edges is defined on a modularly structured network. The bottom panel shows the corresponding transport matrix $\mathbf{E}[A]$, which depends on the network structure given by its adjacency matrix, $A$.}
 \label{fig_T-R_network}
\end{figure}

Here, we present analytical results on how information is transmitted in a complex network by interpreting a subset of its nodes to be transmitters of physical flows, and another subset to be the receivers of these flows (as it is shown in Fig.~\ref{fig_T-R_network}). The information transmitted is derived from these physical flows by encoding them similarly to how it is done when random walks are defined in graphs. In the case of random walkers, the voltage potential at the nodes of a resistor network can be linked to the probability of finding the walker at the node \cite{Tetali_1991,Doyle_2000,Newman_2004,Rubido_2013}; however, a voltage is always defined up to an arbitrary reference value. On the contrary, our approach uses the voltage potential differences and the edge weights (resistance) to define the physical flows (currents), thus, defining invariant physical quantities. Moreover, our formulation allows us to derive exact and approximate expressions for the resultant information flows. Consequently, we are able to interpret directly the information transmission capacity of the network, link them straightforwardly to the network topological invariants, such as the node degree, and bridge the gap between single-channel communication systems and multiple-channel communication systems.

\section{Methods and Model}
The starting point in our approach is to define and to analytically solve the flows in a conservative transport network, namely, solve the Voltage-Flow (VF) problem in a network. The model for our VF problem is that of an Ohmic (passive) circuit \cite{Kirchhoff} with input and output currents, i.e., the definition of a resistor network, $\mathcal{G}$, where each edge current, $f_{ij}$, is linearly related to the potential difference, $\Delta V_{ij}$, between the nodes that the edge connects ($i$ and $j$) and its resistance, $R_{ij}$. Specifically,
\begin{equation}
  f_{ij} = \frac{\Delta V_{ij}}{R_{ij}}.
 \label{eq_Ohm}
\end{equation}
The input currents are set in a node subset, $\mathcal{S}\subset\mathcal{G}$, and the total input leaves the network at another node subset, $\mathcal{T}\subset\mathcal{G}$, corresponding to the network's output. In other words, a total current $I$ enters the circuit at some nodes and the same amount $I$ leaves the circuit at another nodes, thus defining a conservative-flow network (which means that the total current arriving to any node is equal to the total current leaving the node). Consequently, the values that the edge flows, $f_{ij}$, take in Eq.~\eqref{eq_Ohm} depend on the location of the source and sink nodes and the magnitude of the total input flow, namely, $f_{ij}^{(\mathcal{S},\,\mathcal{T})}$. Here, without loss of generality, we define a VF problem with $I = 1$ and set a single-source/single-sink system (for an extension of the problem to multiple sources and sinks, see, for example, \cite{Rubido_2014}), thus, the edge flow is $f_{ij}^{(s,\,t)}$.

In particular, the resistor network is derived from the network structure, hence, it is defined from the set $\mathcal{G} = \{\mathcal{V},\,\mathcal{E}\}$, where $\mathcal{V}$ and $\mathcal{E}$ are the node and edge sets, respectively. We restrict ourselves to symmetric networks by setting the edge weights as \mbox{$W_{ij} \equiv A_{ij}/R_{ij}$}, with $i,\,j = 1,\ldots,\,N$, $N$ being the number of nodes in $\mathcal{V}$, $A_{ij}$ the $ij$-th element of the adjacency matrix, and $R_{ij}$ the edge's resistance. In the cases where the network is unweighed, then $W_{ij} = A_{ij}$. Consequently,
\begin{equation}
  \mathbf{G}\,\vec{V}^{(st)}  = \vec{F}^{(st)}\,,
 \label{eq_V-F_problem}
\end{equation}
where the upper-indexes indicate that the VF problem depends on the location of the source-sink nodes, $s$-$t$, $\vec{F}^{(st)}$ is the flow vector containing the total current at each node (i.e., $[\vec{F}^{(st)}]_i = \sum_j f_{ij}^{(s,\,t)} = \delta_{is} - \delta_{it}$ for $i = 1,\,\ldots,\,N$, $\delta_{ij}$ being the Kronecker delta), and $\mathbf{G}$ is the weighed Laplacian matrix whose entries are
\begin{equation}
 G_{ij} = \left\lbrace \begin{array}{lcl}
                         \sum_{k=1}^N W_{ik} & \text{if} & i = j\,, \\
			  - W_{ij} & \text{if} & i \neq j\,.
                        \end{array} \right.
 \label{eq_laplacian_def}
\end{equation}

The solution for the VF problem [Eq.~\eqref{eq_V-F_problem}] is achieved once the voltages at each node are found from inverting $\mathbf{G}$. However, because $\mathbf{G}$ is a Laplacian matrix, its inverse is ill-defined (its kernel has non-null dimension). Despite of this, we can use the Moore-Penrose pseudo inverse matrix to invert $\mathbf{G}$, allowing to find an exact value for the voltage differences \cite{Rubido_2014,Bollobas,FanChung,Wu_2004,Arpita_2008}. The result is that
\begin{equation}
  \Delta V_{ij}^{(s,\,t)} = \sum_{n = 2}^{N} \frac{ \left( \Delta\left[ \vec{\psi}_n \right]_{ij} \right) }{\lambda_n} \left( \Delta\left[ \vec{\psi}_n \right]_{st} \right),
 \label{eq_voltages}
\end{equation}
where $\vec{\psi}_n$ is the $n$-th eigenvector associated to the eigenvalue $\lambda_n$ of $\mathbf{G}$ (i.e., $\mathbf{G}\,\vec{\psi}_n = \lambda_n\,\vec{\psi}_n\;\forall\,n$) and $\Delta [\vec{\psi}_n]_{ij} \equiv [\vec{\psi}_n]_{i} - [\vec{\psi}_n]_{j}$ is the $i$ and $j$ eigenvector-coordinate difference for the $n$-th eigenmode. We note that the first eigenmode, $n = 1$, corresponds to $\lambda_1 = 0$ and $\vec{\psi}_1 = \vec{1}/\sqrt{N}$, which is the condition that any Laplacian matrix row sum is null, hence, its kernel has at least dimension $1$ \cite{Bollobas,FanChung}. Also, the edge flows, $f_{ij}^{(s,\,t)}$, are straightforwardly derived from this expression by means of Eq.~\eqref{eq_Ohm}. Furthermore, Eq.~\eqref{eq_voltages} shows how the location of the source and sink nodes modify the voltage difference value by means of the corresponding eigenvector coordinate difference.

Equation~\eqref{eq_voltages} is the main derivation that allows us to define our information flows approach. It is applicable to any connected weighted graph and is extendible to many sources and sinks with different inputs and outputs, as long as flow conservation is fulfilled. Specifically, \emph{we define our information flows} from the following flow-transmission probability matrix, $\mathbf{\Pi}^{(s,\,t)}$, with $ij$ entry
\begin{equation}
  \Pi_{ij}^{(s,\,t)} \equiv \frac{1}{d_i} \left| f_{ij}^{(s,\,t)} \right|,
 \label{eq_ProbMat}
\end{equation}
where $d_i \equiv \sum_j^N \left| f_{ij}^{(s,\,t)} \right|$, which is twice the total inflow (by conservation, it also means twice the total outflow) at node $i$ because $f_{ij}^{(s,\,t)} = - f_{ji}^{(s,\,t)}$. By constructing this transition matrix [Eq.~\eqref{eq_ProbMat}], we are detaching ourselves from random walks (RW) on graphs, which solely depend on the network's structure, and we are including the functionality of the network as a communication channel.

A RW on a graph is defined by a transition probability that depends on the graph's structure, namely, $\Pi_{ij}^{(RW)} \equiv W_{ij}/d_i$, where $W_{ij}$ is the $ij$ edge's weight and $d_i$ the weighed degree. This probability matrix is asymmetric ($\Pi_{ij}^{(RW)} \neq \Pi_{ji}^{(RW)}$) but it fulfils detailed balance, i.e., $p_i\, \Pi_{ij}^{(RW)} = p_j\, \Pi_{ji}^{(RW)}$ for every node $i$ and $j$ and probability distribution $\vec{p}$ \cite{Gobel_1974,Lovasz_1993}. Hence, it determines a Markov process: $\vec{p}(n) = \vec{p}(0)\,[\mathbf{\Pi}^{(RW)}]^n$, $\forall n\in\mathbb{N}^+$. Consequently, it converges to a unique stationary probability distribution (SPD), regardless of the initial condition.

\section{Results and Discussion}
Our approach to define information flows (i.e., the flow-transmission probability matrix, $\mathbf{\Pi}^{(s,\,t)}$), allows us to include the network's structure and function into the problem. In other words, our random walker takes into account the network's structure at each node (e.g., how likely is to jump to any of its neighbours given the number of neighbours it has) and also the net current strength at the node ($\sum_j^N | f_{ij}^{(s,\,t)} |$), which is determined by the locations of $s$ and $t$. In this way, we force a direction for the random walker to follow the paths that take it from the source (transmitter) to the sink (receiver). Moreover, since our formulation follows the main mathematical properties of the classical RW on graphs \cite{Gobel_1974,Lovasz_1993}, our $\mathbf{\Pi}^{(s,\,t)}$ converges to a SPD as well, regardless of the initial conditions, namely, regardless of the initial probability distribution (which is independent of $s$ and $t$). Specifically, $\vec{p}(n)^{(s,t)} = \vec{p}(0)\,{\mathbf{\Pi}^{(s,\,t)}}^n \xrightarrow{n} \vec{\mu}^{(s,t)}$, where $\vec{\mu}^{(s,t)}$ and $\vec{p}(n)^{(s,t)}$ inherit the dependence on $s$ and $t$ because of our flow-transition matrix definition in Eq.~\eqref{eq_ProbMat}.

We note that in the RWs on unweighed graphs, the SPD is $\mu_i^{(RW)} = k_i/\sum_i k_i$ \cite{Gobel_1974}, where $k_i = \sum_{j=1}^N A_{ij}$ is the $i$-th node degree. Hence, the RW SPD is directly dependent on the degree distribution of the network alone. On the other hand, our SPD, $\vec{\mu}^{(s,t)}$, depends on where the transmitter and receiver are located as well as on the flow-weighed degree distribution (i.e., it depends on $d_i = \sum_j^N | f_{ij}^{(s,\,t)} |$). Nevertheless, analogously to the RW, our SPD can be derived from our flow-transition probability matrix (FTPM), $\mathbf{\Pi}^{(s,\,t)}$, using standard techniques \cite{Gobel_1974,Lovasz_1993} as we do in what follows.

The reason to obtain a unique SPD is that our FTPM is a stochastic matrix with positive entries that has unit-row-sums ($\sum_j \Pi_{ij}^{(s,\,t)} = 1\;\forall\,i$), as it can be directly verified from Eq.~\eqref{eq_ProbMat}. Consequently, its eigenvalue spectra is bounded in the complex plane by the disc of unit radius, as the Gershgorin theorem predicts \cite{FanChung}. Moreover, this means that there is a non-degenerate maximum-valued eigenvalue $\alpha_1 = 1$, which we arbitrarily set to the first eigenmode of the FTPM, that is non-vanishing. In what follows, we assume that the FTPM is irreducible and aperiodic, thus, $\alpha_1$ is non-degenerate and its right eigenvector set, $\mathbf{P} = \{\vec{v}_1,\ldots,\vec{v}_N\}$, is orthonormal (i.e., $\vec{v}_i \cdot \vec{v}_j = \delta_{ij}\;\forall\,i,j$) \cite{Gobel_1974,Lovasz_1993}. Since, $\mathbf{\Pi}^{(s,\,t)}\,\vec{v}_n = \alpha_n\,\vec{v}_n$ for every eigenmode, the stationary FTPM is
\begin{equation}
  { \mathbf{\Pi}^{(s,\,t)} }^n = \mathbf{P} \mathbf{\Lambda}^n \mathbf{P}^{-1} \xrightarrow{n} \mathbf{P} \left[ \begin{array}{ccc}
						  1 & 0 & \cdots \\
						  0 & 0 & \cdots \\
						  0 & 0 & \ddots
						\end{array}   \right] \mathbf{P}^{-1},
\end{equation}
where $\mathbf{\Lambda}$ is the diagonal matrix containing the eigenvalues of the FTPM (i.e., $\mathbf{\Lambda} = \{1,\alpha_2,\ldots,\alpha_N\}$) and $\mathbf{P}$ is the matrix containing in each column the right eigenvectors of the FTPM, which depend on $s$ and $t$. This means that the limit process ends selecting the product between $\mathbf{P}$'s first column, that is $\vec{v}_1 = \vec{1}/\sqrt{N}$ (as with the Laplacian matrix), and $\mathbf{P}^{-1}$'s first row. Thus, the stationary FTPM entries are given by
\begin{equation}
  \left[ { \mathbf{\Pi}^{(s,\,t)} }^n \right]_{ij} \xrightarrow{n} \frac{ [\mathbf{P}^{-1}]_{1j} }{ \sqrt{N} },\;\;\forall\,i = 1,\ldots,N.
 \label{eq_FTPM_converge}
\end{equation}

The \emph{first result} from taking our approach and define flows on the network, is that we are able to derive an expression for our SPD, $\vec{\mu}^{(s,t)}$, in terms of Eq.~\eqref{eq_FTPM_converge}. Specifically, for any initial distribution $\vec{\mu}_0$ converges to
\begin{equation}
  \vec{\mu}_0\,{ \mathbf{\Pi}^{(s,\,t)} }^n \xrightarrow{n} \left( \frac{ [\mathbf{P}^{-1}]_{11} }{ \sqrt{N} }, \frac{ [\mathbf{P}^{-1}]_{12} }{ \sqrt{N} }, \ldots \right) = \vec{\mu}^{(s,t)}.
 \label{eq_SPD}
\end{equation}
Consequently, we can compute the \emph{amount of information encoded in the network}, $H^{(s,\,t)}$, due to the flows in the particular $s$-$t$ configuration by
\begin{equation}
  H^{(s,\,t)} = - \sum_{i = 1}^N \mu_i^{(s,\,t)} \log\left[ \mu_i^{(s,\,t)} \right].  
 \label{eq_st_Entropy}
\end{equation}
We note that the expression for $\vec{\mu}^{(s,\,t)}$ is also the one for the left eigenvector of the FTPM maximum-valued eigenvalue, because the Markov process defined by the FTPM converges when $\vec{\mu}^{(s,t)} = \vec{\mu}^{(s,t)}\,\mathbf{\Pi}^{(s,\,t)}$. As a left eigenvector, it has to have unit $L_1$-norm, i.e., $\left\| \vec{\mu}^{(s,\,t)} \right\|_1 = 1 = \sum_i \left|\mu_i^{(s,\,t)}\right|$, because they constitute probability distributions, contrary to the right eigenvectors of $\mathbf{\Pi}^{(s,\,t)}$ which have unit $L_2$-norm (also known as the Euclidean norm).

The network's maximum information, namely, its information capacity $C$, is achieved when searching for the maximum $H^{(s,\,t)}$ after changing $s$ and $t$'s location around the network, i.e.,
\begin{equation}
   C \equiv \max_{ s,t\in\mathcal{G} }\left\lbrace H^{(s,\,t)} \right\rbrace.
 \label{eq_InfCapacity}
\end{equation}
Because we have an analytical expression for the flows in the network [Eq.~\eqref{eq_voltages}], we can compute the information flows directly [Eq.~\eqref{eq_ProbMat}], thus, we are presenting an efficient approach to compute $C$ straightforwardly.

The \emph{second result} comes from the information shared in the network between source and sink at any given time $n$, i.e., the mutual information (MI). In particular, we find that under the SPD conditions, the network's MI is null. In general, the mutual information of a given $s$-$t$ configuration at an instant $n$ is defined as
\begin{equation}
  I_{(s,\,t)}(n) \equiv \sum_{i,j}^N p_{i,j}^{(s,\,t)}(n) \log\left[ \frac{ p_{i,j}^{(s,\,t)}(n) }{ p_i(n)^{(s,\,t)}\,p_j(n)^{(s,\,t)} } \right],
 \label{eq_MutInf}
\end{equation}
where $p_{i,j}^{(s,\,t)}(n)$ is the joint probability to find node $i$ and $j$ having the same event at time $n$, which in our case, corresponds to the same information flows determined by Eq.~\eqref{eq_ProbMat}. In order to understand this, we recall that a joint probability can always be defined from the probability distribution and the transition probabilities by
\begin{equation}
  p_{i,j}^{(s,\,t)}(n) \equiv p_i(n)^{(s,\,t)}\,\left[ \mathbf{\Pi}^{(s,\,t)} \right]_{ij}.
 \label{eq_JointProb}
\end{equation}
This expression also shows that the joint probability always depends on the location of $s$ and $t$, contrary to the initial probabilities $p_i(0)$. According to Eq.~\eqref{eq_JointProb}, when $n \to \infty$ we retrieve the SPD, $p_i(n) \mapsto \mu_i^{(s,t)}$, and the stationary transition probability, $[\mathbf{\Pi}^{(s,\,t)}]_{ij} \mapsto \mu_j^{(s,\,t)}$ [Eq.~\eqref{eq_FTPM_converge}]. Thus, the joint probability becomes decoupled. In other words, when the information flows become stationary, the probabilities at each node are independent quantities which only depend on the location of the source and sink and the network structure.

As a \emph{final result}, and to illustrate our approach, we apply Eqs.~\eqref{eq_ProbMat}-\eqref{eq_st_Entropy} to an unweighed modular network with $N = 100$ nodes, two identical modules with $50$ nodes each, identical edge density $\rho = 0.5$, and Erd{\"o}s-R{\'e}nyi (ER) network characteristics \cite{Erdos}. We place $10$ extra links randomly between the modules to inter-connect them.

\begin{figure}[htbp]
 \begin{center}
  \includegraphics[width=0.8\columnwidth]{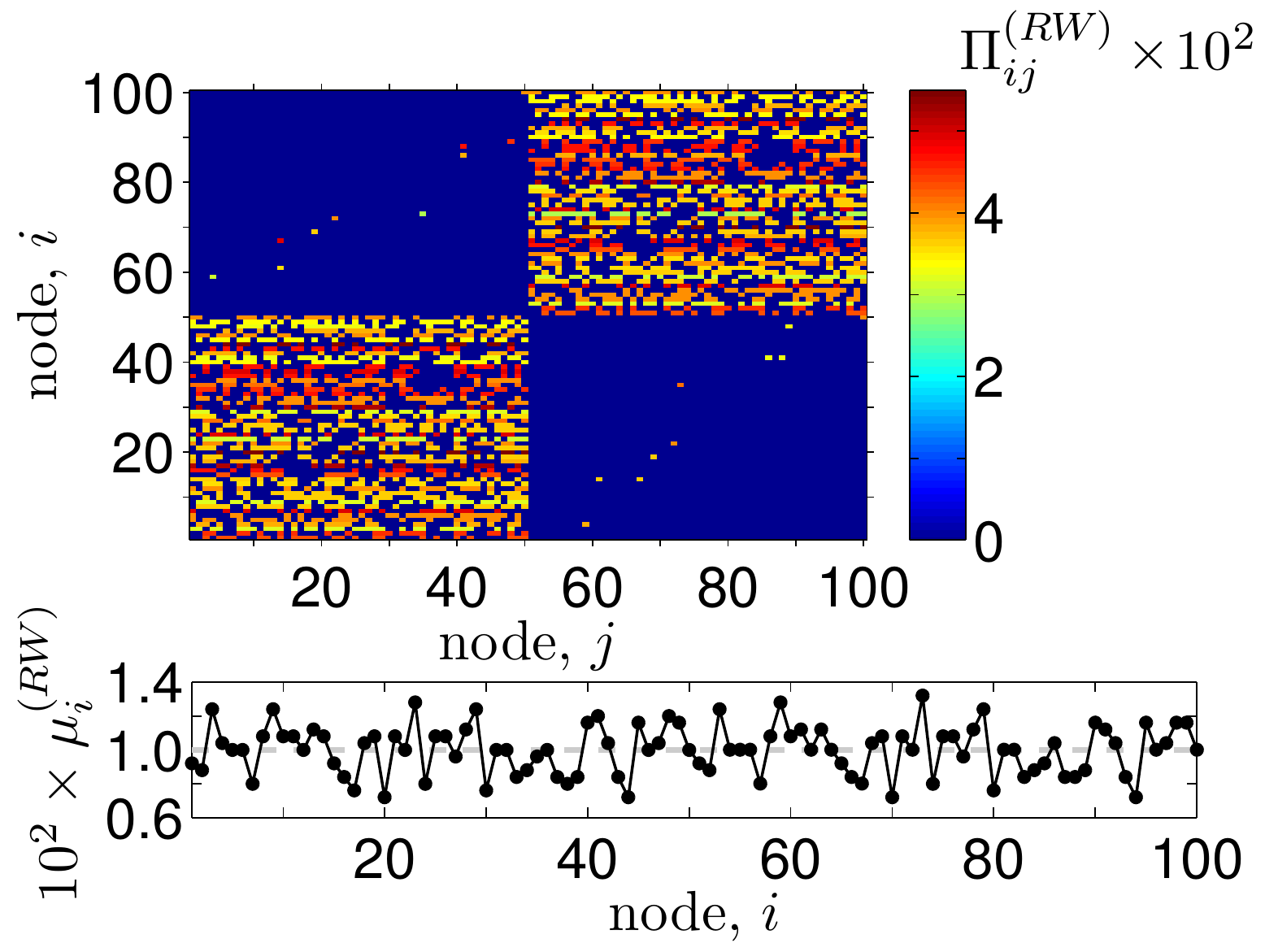}
 \end{center} \vspace{-1pc}
  \caption{Transition probability matrix (TPM), $\mathbf{\Pi}^{(RW)}$ [top panel], and stationary probability distribution (SPD), $\vec{\mu}^{(RW)}$ [bottom panel], for a random walk in a two-module network with $N = 10^2$ nodes. The two modules are identical, have an Erd{\"o}s-R{\'e}nyi topology (namely, they are random) with $50$ nodes and a $0.5$ edge density, and are inter-connected by $10$ edges. The TPM entries are represented by the colour-code. Both, TPM and SPD, are multiplied by $N$ to improve clarity.}
 \label{fig_RW}
\end{figure}

In order to contrast our results with that of a classical RW, we first show in Fig.~\ref{fig_RW} the RW transition probability matrix, $\mathbf{\Pi}^{(RW)}$, (top panel) and the corresponding SPD, $\vec{\mu}^{(RW)}$ (bottom panel). From the top panel, it can be seen that the walker is able to transit through the network only according to the network's structure, as each transition is defined by $\Pi_{ij}^{(RW)} \equiv A_{ij}/k_i$. Moreover, the SPD is the vector of node degrees, $\vec{k}$, divided by the total edge number, which in this case holds $\vec{\mu}^{(RW)} = \vec{k}/2492$. We remind that the expected edge number, $E[M]$, for an ER network of size $N$ is: $E[M] = \rho\,N\,(N-1)/2$, where in our case corresponds to $E[M] = 612.5$ for each module. Thus, taking into account both modules plus the extra inter-links, the total expected edges are $2(2M)+10 = 2460$, which is close to the exact value in our case of study, namely, $2492 = \sum_i k_i$.

\begin{figure}[htbp]
 \begin{center}
  \includegraphics[width=0.8\columnwidth]{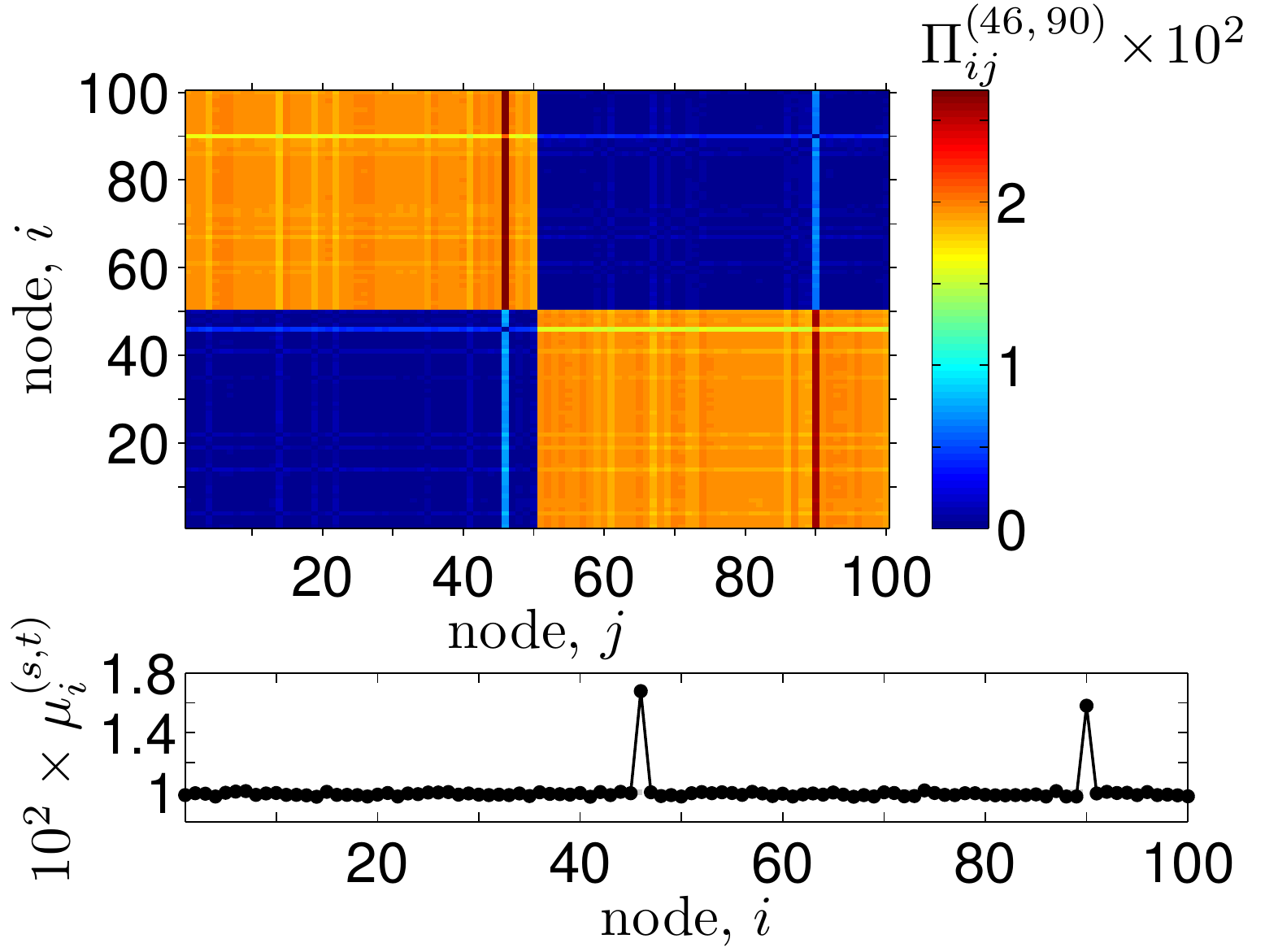}
 \end{center} \vspace{-1pc}
  \caption{Flow-transition probability matrix (FTPM), $\mathbf{\Pi}^{(s,t)}$ [top panel], and stationary probability distribution (SPD), $\vec{\mu}^{(s,t)}$ [bottom panel], for a source, $s$, sink, $t$, configuration in the same network as Fig.~\ref{fig_RW}. Colour codes and normalisation follow the same criteria as in Fig.~\ref{fig_RW}.}
 \label{fig_VF}
\end{figure}

On the other hand, we set a source (transmitter) in one module, $s = 46$, and a sink (receiver), $t = 90$, in the other module. This defines a voltage difference between every node in the network [Eq.~\eqref{eq_voltages}]. Hence, the information flows have a transition matrix that allows the walker to jump freely between modules, as it is seen in Fig.~\ref{fig_VF}. Moreover, we can see that the probabilities to jump between modules are higher than the inner jumps, which is the opposite situation than in the RW. Also, we can directly distinguish the source and sink nodes importance in the FTPM and in the SPD, because both have higher values than the rest of the nodes.

For this $s$-$t$ flow configuration, the information generated is comparable with the RW, as we verify with Eq.~\eqref{eq_st_Entropy}, holding $H^{(RW)} \simeq 6.63$ bits and $H^{(46,\,90)} \simeq 6.64$ bits. However, if we place transmitter and receiver in the same module, most FTPM values become nearly null, with the exception of those that correspond to the paths in the module joining the source with the sink. Consequently, the information decreases. For example, the configuration with $s = 46$ and $t = 30$ results in $H^{(46,\,30)} \simeq 5.74$ bits. This means that the network's information capacity [Eq.~\eqref{eq_InfCapacity}] is mainly influenced by the number of modules in the network, with the maximum achieved when source and sink are set the furthest apart. Similar results are expected for multi-layered networks, since each layer can be thought as a different module.

\section{Conclusions}
In this work we present a rather novel approach to tackle the problem of information transmission in networks. We define a circuit-like circuit based on the network's structure, set a source [sink] node that inputs [outputs] a constant flow, and solve the voltage-flow problem for every edge in the network. By doing this, we are able to define a flow-transition probability matrix, similarly to how it is done with random walks in graphs, but instead of using solely the network's structure, we use the resultant flows. The reason behind defining the source-sink nodes, is that it allows us to interpret the problem as a transmitter-receiver problem whose communication channel has multiple paths. Consequently, with our approach, we have a tractable way to understand how the information is transmitted across a network.

In particular, we see that modular networks have an information capacity that increases with its modularity, namely, with the number of communities in the network. In general, we see that the furthest apart transmitter and receiver are, the higher the information that is generated in the network. This conclusion leads us to think that modular networks have higher information capacities than non-modular networks due to the bottle-necks that inter-links constitute for the flows. However, we note that further research in this line is needed in order to have sound conclusions. Moreover, we expect similar results to hold also for multi-layered or multiplex networks, where source and sinks should be placed in different layers in order to increase the capacity.

\section{Acknowledgements}
NR acknowledges the support of PEDECIBA, Uruguay. CG and MSB thank the Scottish University Physics Alliance (SUPA) support. MSB also acknowledges the support of EPSRC grant Ref. EP/I032606/1.


\end{document}